\documentclass[11pt,a4paper]{article}
\usepackage[dutch, english]{babel}
\usepackage[utf8]{inputenc}
\usepackage[T1]{fontenc}
\usepackage[left=2.5cm, right=2cm, top=2cm]{geometry}
\usepackage{amsmath}
\usepackage{amsfonts}
\usepackage{amssymb}
\usepackage{graphicx}
\usepackage{caption}
\DeclareCaptionFont{mdit}{\mdseries\itshape}
\usepackage{subcaption}
\usepackage{mathptmx}
\usepackage{multirow,tabularx}

\usepackage{hyperref}
\usepackage{hyphenat} 
\usepackage[natbibapa]{apacite}
\begin{document}
	
\selectlanguage{english}

\bigskip

\part*{\centering{`Corona-safe' Measures for Cyclists at Intersections}}

\bigskip

\begin{center}
	A. Maria Salomons 	\\ 
	Delft University of Technology, Transport en Planning 	\\ 
	e-mail: a.m.salomons@tudelft.nl 	\\ 
	phone:  +31 615235756 	\\
	
\end{center}

\medskip

\subsubsection*{ Abstract}

During the Corona crisis measures were taken to avoid the spread of the virus, and one of the most important was to `keep sufficient distance'. In this period cyclists were facilitated in distancing by the widening of cycle paths, or the use of car lanes. The Corona measures also affected intersection control, since one of the principles of signalizing is based on clustering of traffic. The settings of the control take into account the way cyclists tend to cluster close together at the stop line, so for optimal control, if the clustering changes, the control should be adapted as well. This paper presents measures for `Corona-safe' intersection control for cyclists, and discusses the effect of the measures and their (dis)advantages. Further it was investigated how some municipalities used these measures and conveyed about it in the media and how these measures were received by the public.

The measures can be divided in two groups: detection and timing. For the municipalities considered, the least popular measures concerned the detection methods, such as adaptations in use, or type, of push button, since the negative aspects can be large (effect on throughput and costs). More popular were changes in the timing of the control, the most conveyed measures being more frequent and longer green for cyclists.

The acceptance of the measures by the public (determined via reactions on social media) is mixed, some are positive about the improved control, others negative (effect on car traffic or costs). The most common reactions are general complaints about intersection control, and about cyclists behaviour. 

For future use the measure `more frequent green'  is most suitable, permitted cyclists are detected to support the acceptance of the control. Also bicycle apps can be applied, since these results in more comfort and have limited negative effects.

\subsubsection*{Keywords}
Corona~crisis, Cyclists, Intersection~control  

\newpage

\section*{Introduction}

At the end of February 2020 the Corona pandemic appeared in the Netherlands, and in March 2020 it became clear that to be sure the care would be able to handle the number of seriously ill patients, measures had to be taken. In the Netherlands the Rijksinstituut voor Volksgezondheid en Milieu (RiVM) advised the Dutch government to restrict the number of contacts between patients, leading to the decision of an `intelligent lockdown', which came into effect of March 12, 2020  (\cite{GN}). For this lock-down it was advised to stay inside as much as possible, but it was not forbidden to go outside. When going outside, a distance of 1.5 m was enforced  (\cite{VB}, \cite{GN1}). The restrictions had a large influence on traffic volumes  for both public transport (\cite{Hor}), car traffic and bicycle traffic (\cite{EDP}). This paper aims for a description of (possible) measures that were taken during the pandemic, which measures were communicated to the public and the reactions of the public, and will consider the effectiveness and the (dis)advantages of the measures.

Based on these findings, a recommendation is given of the use of these measures, as well as for an ongoing pandemic, or as a comfort measure for cyclists in the future. The measures can also be found in \cite{GNMI} where section 4.5 is partly based on the findings described in this paper. A presentation of this paper can be found at the weblog of the author, \cite{Sal}.

\section*{Research goal}
Since it was not forbidden to go outside, measures were taken to keep travellers safe while travelling, for instance  public transport the number of seats were reduced  (\cite{Hor}, \cite{NOSa}). Also for pedestrians and cyclists measures were taken to make the `one and a half meter society' possible, such as widening of cycle paths (\cite{FB}, \cite{Sus}). The goal of this research is to  describe which measures  were considered and/or taken by municipalities to facilitate cyclists at signalized intersections. The control of these intersections was optimized for cyclists considering the Corona situation, with the assumed effect on avoiding contamination.

Another goal is to determine which measures are beneficial for cycling, which can also be considered as a comfort measure for cyclists in normal situations, when the Corona pandemic is contained.

\section*{Research method}
The research method is a literature study and a survey (online and by means of personal communication) among engineers and consultants responsible for traffic control at intersections. The literature contained mainly `grey papers' as governmental statements, news papers and news sites, social media as Facebook and LinkedIn. Possible measures are listed, which measures were chosen by municipalities and communicated to the public. In the online survey and the personal communication traffic engineers were asked which measures were applied in their municipality or province, whether these measures were communicated with the public and if these measures will be applied in future as comfort measure in future. Further, reactions of the public (on social media where the measures were explained) were collected to determine the acceptance of the measures. 

\section*{Intersection control and Corona measures}
In this section is described how control at intersections is realised in the Netherlands, and what extra measures were advised, considered and applied.

\subsection*{Intersection control and control measures}
Next to safety, one of the important aims of intersection control is to optimize the throughput of an intersection. In the Netherlands this is done by vehicle-actuated control, where vehicles on directions are clustered, and a direction given green for at least the minimum green time, and after that extended  while the queue lasts, or until the maximum green time is reached. The control structure, which determines the order and combinations in which the directions are given green, remains the same, but the green time of a direction varies from cycle to cycle. If there is no traffic present, a direction can be skipped in the structure, leading to shorter cycle times, with adequate green times (\cite{Mul}, \cite{Fur}).  The control structure is determined by optimizing the control for the average demand, resulting in a low minimum cycle time. This control structure can be changed by time of day as the average demand patterns change. This method of control is applied to all traffic modes, so car traffic as for cyclists and pedestrians, and for the control structure determination, all traffic directions are taken into account. 

When vehicle actuated control is used, correct detection is important to grant that sufficient green time is given to the traffic flow. For car traffic and cyclists, this is mostly done by means of induction loops (\cite{Ste}). For car traffic normally two or more detectors are installed, one close to the stop line, and one or more further upstream. For cycle crossings one detector is placed close to the stop line, less frequent there is a second detector 20 m upstream. Also push buttons are placed at bicycle signals, to avoid that bicycles with a low amount of magnetic material are not detected (e.g. carbon bikes). 

\subsubsection*{Push buttons}
Given the Corona situation, the safety of the push button was discussed, since at the beginning at the Corona crisis it was unclear how long the virus stays active on surfaces (\cite{NIH}). While in other countries this question was important for pedestrian crossings( \cite{Cou}), in the Netherlands this discussion also extended to cyclists, with this difference that if there are induction loops present, often it is not needed to use the push button to be detected. Therefore in some cities, like Rotterdam( \cite{Koo}) and Apeldoorn (\cite{IdBA}) messages were applied that it is not necessary to use the push button. If this is applied for pedestrians directions, that are not detected  to avoid these directions do not get green, a `cyclic demand' is needed where every cycle the pedestrian direction will turn green. This causes longer cycle times with unused green where other traffic has to wait `for nothing'. To avoid this situation, municipalities decided on quicken the renewal of push buttons, where old push buttons that are only to be activated using a finger (a  sunk push button), by push buttons that can be activated by the elbow or foot (\cite{Schie}).

\subsubsection*{Detectors}
For induction loops there are also Corona related issues. One of the issues appeared at the very beginning of the Corona crisis when the traffic volumes were very low. If induction loops show much less spikes than in a normal situation on a comparable day, the detector is considered to have a failure and is switched off automatically (D.Poncin, personal communication, March 31st, 2020). Given this extreme circumstances, for all directions concerned, cyclic demand can be installed.

If the detectors are functioning, also correct determination of the flow of cyclists is affected by the 1.5 m measure.The time a detector is not occupied during the green time is measured, and if the detector is not occupied for at least the so-called gap-out time, the flow over the stop line is considered to have ended, and the green time for this direction is ended, provided the green time has lasted at last the minimum green time. If cyclist have to take 1.5 m distance, the queue will solve more slowly, and it will take more time for cyclist to reach the detector from the back of the queue, especially when there is only one detector. Under normal circumstances, cyclist are close to each other near the stop line (20-30 cyclist between stop line and second detector 20 m upstream, (\cite{Reg})). Given that the distances are larger and the queue takes longer to solve, the minimum green time (\cite{CRO1} )and the gap time (\cite{DTV} ) should be enlarged.

\subsubsection*{Green realisation and duration}
Even though the green time is dependent on the detection  it is beneficial for the throughpunt of cyclists and the queue reduction to increase the minimum green time, especially when only one detector is present. Multiple realisations of green for cyclists during one cycle will also reduce the queue and the delay. The municipality of Leiden uses ``wait-in-green''  for cyclist, in this case the traffic signals of bicycle directions remain green until conflicting traffic is detected. When compared to ``wait-in-red'' for all directions, this leads to less unnecessary stops for the cyclists, but traffic at the conflicting direction has to decelerate more since it takes some time to end the green, apply the yellow time and clearance time. 

Where normally the control is designed as conflict-free, the waiting time of cyclist will be higher than when conflicts between turning car directions are permitted (permitted conflicts). Permitted conflicts increase the efficiency (\cite{Mach}), but it can impair the safety especially when traffic participants are not aware conflicts can occur when normally the crossing is conflict-free.

\subsection*{Results: proposed measures}
In Table \ref{tab:proposedMeasures} the measures are listed that were proposed by the Fietsersbond (the cyclists unions)( \cite{FB}), the consultancy \cite{DTV}, and proposed by municipalities (found in the media or via the survey). For all measures the advantage for the cyclists is listed, and disadvantages. The disadvantages can be costs, but also the traffic safety can be impaired. Measure 6 is an example  that can lead to unsafe situations: if cyclists suddenly start drive on car lanes, the motorists are not used to come across cyclist and this unexpected situation (even when clearly indicated with signs) will lead to dangerous situations. In Amsterdam at various busy locations cyclists were directed to the car lane to give more space to both cyclists and pedestrians, but at the Rijnstraat it appeared to be too dangerous (\cite{Ams}).  Since the car demand is high in this street, the queues at the intersections are large. It appeared it was not possible for cyclists to pass the queue to the stop line and cyclist had to wait in between the cars, leading to an even longer queue, which made car drives to move over to the tram lane. Therefore on this particular location, this measure was reversed again. 

One of the measures is take bicycle priority at iVRIs into account, iVRI stand for the Dutch abbreviation of intelligent traffic control (\cite{iVRI}). iVRI controllers communicate with the surrounding traffic, so traffic can be detected at an earlier moment than with detector loops. If the route of the traffic participant is conveyed to the control, the control can anticipate on the routes of the traffic. A bicycle app can learn the the preferred route from its user (\cite{Ver}), in this case the route recognition is part of the mobile phone of the user, rather than part of the controller, for the controller the app will act as a detector.

\begin{table}[h]
\caption{Proposed measures to facilitate cyclists directions at signalised intersections; proposed by municipalities (via media (M), survey of personal communication(S)), Fietsersbond (FB)  and DTV }
	\begin{tabular}{|c|l|l|l|l|}
		\hline
		\textbf{\#}  & \textbf{By} & \textbf{Measure} & \textbf{Advantage} & \textbf{Disadvantage} \\
		& &\textbf{for cyclists} &\textbf{for cyclists}&\\
		\hline
		\hline
		1&M&No push buttons and &Hygienic&Increased delay\\
		&&cyclic demand&&\\	
		\hline	
		2&M& Other types of push buttons &Hygienic&Costs\\
		\hline
		3&S&Infra-red cameras &Hygienic, sufficient green&Costs\\	
		\hline	
		4&S&Monitor the traffic &Less delay/stops&\\
		&& take action if needed&&\\
		\hline	
		5 & FB& Turn off all traffic signals & Less delay/stops & Only possible for\\
		&&&&low traffic volumes\\
		\hline
		6 & FB& Cyclists on car lanes & More space  & Cyclist might not be \\
		&& ($\le$30 km/h)&&detected\\
		\hline
		7 & DTV & Shorter green car directions & Less red time  & Risk: head-tail collisions \\
		&&&& in moving car queue\\
		\hline
		8 & DTV & Less priority for PT & Less red time  & Negative effect on PT  \\
		&&&&running on time\\
		\hline
		9 & DTV & Multiple green per cycle & Smaller queues  & Longer car-cycle times \\
		\hline
		10 & DTV & Permitted conflicts & Less red time  & Unsafe due to\\
		&&&& new uncontrolled conflict \\
		\hline
		11 & DTV & All cyclist green at the same time & Less stops  & Bicyce-bicycle conflicts \\
		\hline
		12 & DTV & Coordination of bicycle directions & Less stops  & Longer car-cycle times \\
		\hline
		13 & DTV &increase priority for cyclists & Less red time & Longer car-cycle times \\
		\hline
		14 & DTV & Increase min./max.green & Longer green  & Unused green \\
		\hline
		15 & DTV & Increase detector gap-out time & Longer green & Unused green \\
		\hline
		16 & DTV & Bicycle detection apps & Less stops & Longer car-cycle time \\
		\hline
		17&S&Wait-in-green for cyclists& No unnecessary stops & Deceleration conflicts\\
		\hline
		18&S&Take away green wave for cars&Reduced delay&Multiple stops for cars\\
		\hline
	\end{tabular}
	\label{tab:proposedMeasures}
\end{table}

\subsection*{Results: which measures were applied by municipalities}

In Table \ref{tab:measuresMunicipalities} the municipalities are listed from small to large city, with the measures taken in this city, and how this information was gathered (via media\footnote{For PDF: in Table \ref{tab:measuresMunicipalities} the names of the municipalities are hyperlinks to the websites.} or via survey).  The date indicates the date the measures were made public, or the date the survey was filled in.

One city, Eindhoven, clearly indicated that no measures were taken and this was also not conveyed to the public. The reason given was that the virus did not seem to be very contagious in the open air, so the risk for pedestrians and cyclist can stand close together at intersections is very low. Also the Province of Gelderland did not apply any measures, because they considered their 160 intersections `corona proof', with suitable detector- and green time settings.

The municipality of Delft indicated that next to the measures taken now, in future prioritizing cyclists at intersections with either bicycle detection apps, intelligent traffic signals, and more green per cycle for cyclists are considered, this independent from the pandemic.

\begin{table*}
\caption{Municipality (-hyperlink),  number of inhabitants, Measures (from table \ref{tab:proposedMeasures}), Method: research method to determine which measures were applied, by media  (M) or by survey or personal communication (S), dd: the date the measures were published or communicated with the author of this paper, Reactions: number of likes and written positive reactions (pos.), dislikes and negative (neg.) written reactions, and other (oth.) reactions.}  
	\begin{tabular}{|l|l|l|l|l|ll|ll||l|}
		\hline
		{\textbf{Municipality/} }& \textbf{\#inh.} & \textbf{Mesaures} & \textbf{Method} & \textbf{dd}&\multicolumn{5}{c|}{\textbf{Reactions}} \\
		{\textbf{Province} }& $x10^5$  & & &\textbf{2020} &\includegraphics[width=0.05\linewidth]{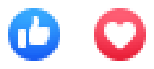} &{\footnotesize \textbf{pos.}}& \includegraphics[width=0.05\linewidth]{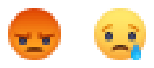}& {\footnotesize \textbf{neg.}}& {\footnotesize \textbf{oth.}}\\
		
		\hline
		\hline
		\href{https://www.ad.nl/gouda/groen-licht-voor-fietsers-zij-kunnen-binnenkort-sneller-het-kleiwegplein-oversteken~a75aca38/}{Gouda}& 	0.7	 &9, 14 &M& 28-05&&&&&\\
		\hline 
		\href{https://www.facebook.com/gemeenteschiedam/posts/2886901181357498}{Schiedam}&  	0.8 &	2 &M&28-05&183 &4&1& 14&0\\
		\hline
		\href{https://www.facebook.com/gemeentedelft/posts/1483430378523565}{Delft} &	1.0	 &9, 14, 15, 16 &M, S& 18-06&109& 7& 3& 1&6\\
		\hline
		\href{https://www.destentor.nl/deventer/fietsen-door-deventer-om-nieuwe-app-te-testen-te-ingewikkeld-geef-auto-s-standaard-rood-en-fietsers-groen~abd0bf98/}{Deventer}  &	1.0	 &16&M&16-06&&&&&\\
		\hline
		\href{https://www.facebook.com/ADZoetermeer/posts/fietsen-geldt-als-een-van-de-veiligste-manieren-van-vervoer-in-tijden-van-corona/2639375579671677/}{Zoetermeer} &	1.2	 &2 &M&  21-0&13 & 2&4& 10&\\
		\hline
		Leiden &	1.3	 &1, 9, 17, 18 &S& 10-08&&&&&\\
		\hline	
		\href{https://www.facebook.com/gemeente.haarlem/posts/3031638510231367}{Haarlem} &	1.5	  &7, 9 &M&   11-05&300& 6& 0& 4&\\
		\hline
		\href{https://www.facebook.com/gemeenteenschede/posts/3110744732318581}{Enschede}  &	1.6	 &4, 9, 14 &M& 03-06&16& 2 &0& 0 &7\\
		\hline
		\href{https://gemeente.groningen.nl/actueel/nieuws/coronamaatregelen-voor-een-veilige-en-bereikbare-binnenstad}{Groningen } &	2.0	 &5, 9, 11&M&  28-05&&&&& \\
		\hline
		\href{https://www.utrecht.nl/bestuur-en-organisatie/coronavirus/maatregelen-coronavirus/meer-ruimte-voor-voetgangers-en-fietsers/}{Utrecht} &	3.5	 &4 &M& 28-06? &&&&&\\
		\hline
		Eindhoven&  3.7 &none&S&02-07&&&&&\\
		\hline
		\href{https://www.ad.nl/den-haag/verkeersregels-in-coronatijd-verkeerslichten-bij-veertig-kruispunten-voor-fietsers-langer-op-groen~ae108d0c/}{The Hague} & 5.4 & 4, 7, 14, 15 &M& 11-05&&&&& \\
		\hline	
		\href{https://indebuurt.nl/rotterdam/gemeente/wen-er-maar-aan-eenrichtingsverkeer-op-de-stoep-en-minder-parkeerplekken~130119/}{Rotterdam}  &	5.8	 &1, 14 &M, S& 28-05&&&&&\\
		\hline
		\href{https://www.facebook.com/gemeenteamsterdam/posts/4147294781978404}{Amsterdam} &	8.6 &	3, 4, 6, 7, 12, 14 &M&  27-05&233 & 2&22&5&80\\
		\hline
		\hline
		{Prov. North Holland }&37.0&4, 13 &S& 24-08&&&&&\\
		\hline
		{Prov. Gelderland}    & 20.1 &none &S& 09-07&&&&&\\
		\hline
	\end{tabular}
	\label{tab:measuresMunicipalities}
\end{table*}
\subsection*{Reactions of the public}
From social media website the reactions of the public were collected, to determine the acceptance of the measures. The reactions are both the number of likes
\includegraphics[width=0.05\linewidth]{Like} and dislikes \includegraphics[width=0.05\linewidth]{Dislike}, and written positive and negative reactions, and other written reactions that are neither positive or negative about the applied measures. The written reactions can be biased, because on social media more people tend to react negatively than positively (\cite{Ros}). The reactions can give insight in how the comfort of these measures are experienced, which can be of importance for future use of the measures. 

Overall, the number of likes (854) outnumber by far the number of dislikes (30). For the written reactions, Schiedam and Zoetermeer which considered new push buttons, received the most negative written reactions (24 negative, 10 positive) in the sense of 'waste of money'. For Delft, Haarlem, Enschede and Amsterdam the written positive reactions on the control measures (longer, more often green) were slightly in the majority (17 positive, 15 negative). For Delft, Haarlem, and Enschede the`other reactions' mostly comprised remarks and questions about certain intersections, where the control was described as being sub-optimum for cyclists. For Amsterdam a large majority of the reactions contained negative reaction about cyclists behaviour (48 reactions of the 87 reactions in total).

\section*{Conclusions and Recommendations}
Several cities in the Netherlands have taken measures, where `multiple green per cycle' and `increase min./max green time' were mentioned most often. Two measures were not mentioned, `less priority for PT' and `permitted conflicts'. The PT-priority was not affected because municipalities attach importance to public transport and public transport was also reduced during the intelligent lockdown (\cite{Hor}). The `permitted conflicts' were deemed too dangerous with respect to the gain of lower odds of contamination.

The effect of the measures in the battle against the spreading of the Corona virus is debatable. The city of Eindhoven did not take any measures, since Corona did not seem contagious in the open air, nor for the buttons \cite{Gol}, nor for groups of cyclist which group at the intersection in the open air for a short time (\cite{Jon} ) (a few minutes, waiting times seldom exceed two minutes (\cite{Mul})) .

Nevertheless, some measures also increase the comfort for cyclists, since more space is available and bicycle queues are smaller and the delay at intersections is less. Monitoring intersections during the day can optimize both the delay for cyclists and cars, since measures in favour of cyclist, which lead to delay for cars, will only be applied if necessary. Therefore, in the survey and by personal communication was asked whether these measures would be continued in future, and several respondents (Delft, Deventer, Leiden, Enschede, The Hague, Province of North Holland) indicated that they were willing to do so, especially the bicycle app in combination with iVRI was mentioned.  In both The Hague (\cite{Gor}) as the Province of North Holland (\cite{PNH}) the bicycle monitor will be used in future, independent from the Corona pandemic. Two municipalities explicitly mentioned speeding up the plans to facilitate cyclists better, The Hague developed the bicycle monitor faster, and Leiden used plans from a future policy program.

The recommendation is that measures which reduce the delay for cyclists and reduce unnecessary stops, are suitable in case cyclists are contagious, and in future when the Corona crisis is solved and the bicycle demand is increased: the aim of the Dutch government is 20\% more cyclists in 2027 (\cite{GN2}). 

The reactions of the public were positive in `likes', but critical in written comments, especially about replacing buttons (`waste of money'). The comments about control measures were more positive, and critical in the sense that intersections were indicated where the control can be optimized, so it is recommended to monitor intersections and  improve the control further, if needed. 

In Amsterdam the acceptance for measures that benefit cyclists is impaired by the negative opinion of cyclist behaviour. It is advisable for the municipality of Amsterdam to study how different modes can use the scarce space, to increase mutual  understanding, for a more solid acceptance of measures that facilitate active modes.

\section*{Further research}
In Corona times at various places in the world new bicycle lanes popped up, and cycle usage is increased (\cite{Phy}), but this new cycle lanes will inevitably lead to interactions with traffic at intersections. This can lead to an increase of unsafe situations, as in Berlin where two third of the accidents where cyclists were involved occur at places where traffic interacts (\cite{Gua}).  How do cities cope with these new interactions and what are the developments of traffic control for cyclists? Future research will focus on problems cities face with the increased number of cyclists at intersections, and solutions that are applied. The aim for the research is that next to a sustainable way of transportation, cycling will be healthy and safe.

\bibliography{MeasuresCyclistsIntersectionsCorona}	
	
\end{document}